\documentstyle[aps,epsfig]{revtex}

\begin{document}
%\preprint{MPG ...., UNITU-THEP-24/2001}
\twocolumn
\title{A kinetic approach to $\eta'$ production from a CP-odd phase\vspace*{0.4em}}
\author{D.B. Blaschke}
\address{Fachbereich Physik, Universit\"at Rostock, Universit\"atsplatz 1,
D-18051 Rostock, Germany\\
and\\
Bogoliubov Laboratory for Theoretical Physics, Joint Institute for Nuclear 
Research, 141980 Dubna, Russia\\[0.6\baselineskip]}
%\and
\author{F.M. Saradzhev}
\address{Institute of Physics, National Academy of Sciences of Azerbaijan,
H. Javid pr. 33, 370143 Baku, Azerbaijan\\[0.6\baselineskip]}
%\and
\author{S.M. Schmidt and D.V. Vinnik}
\address{Institut f\"ur Theoretische Physik, Auf der Morgenstelle 14, Universit\"at T\"ubingen,
D-72076 T\"ubingen, Germany\\[0.6\baselineskip]
\parbox{140mm}{\rm \hspace*{1.0em} 
{\rm
The production of $(\eta,\eta')$-mesons during the decay of a CP-odd
phase is studied within an evolution operator approach.
We derive a quantum kinetic equation starting from 
the Witten-DiVecchia-Veneziano Lagrangian for pseudoscalar mesons
containing a $U_A(1)$ symmetry breaking term. The non-linear 
vacuum mean field for the flavour singlet pseudoscalar meson 
is treated as a classical, self-interacting
background field with fluctuations assumed to be small.
The numerical solution provides the time evolution of
momentum distribution function of produced $\eta'$- mesons after a
quench at the  deconfinement phase transition.
We show that the time evolution of the momentum distribution of the
produced mesons depend strongly 
on the shape of the effective potential at the end of the quench,
exhibiting either parametric or tachyonic resonances. Quantum
statistical effects are essential and lead to a pronounced Bose enhancement of the low
momentum states. \\[0.4\baselineskip]
% \\[0.4\baselineskip] 
Pacs Numbers: 25.75.Dw,12.38.Mh,05.20.Dd,05.60.Gg}
}}

\maketitle

\section{Introduction}
Construction of the  Relativistic Heavy Ion Collider (RHIC) 
at the Brookhaven National Laboratory is completed and it is designed to
initiate energy densities sufficient to produce a quark gluon plasma
(QGP) \cite{QM2000}. Such a strongly correlated state of matter has a life time
smaller than $1fm/c$ and due to rapid collisions the plasma
thermalizes, and at critical values of temperature and density the
quarks and gluons form hadronic bound states: a process driven by confinement and
chiral symmetry breaking. Many aspects of the plasma's production
and evolution are characterised by non-linear dynamics. The
hadronisation process itself as well as critical phenomena in the vicinity
of the phase boundary require a study with non-equilibrium techniques.

One challenging example of a far from  equilibrium process
is spontaneous particle  creation in a strong background electric 
field, i.e. the Schwinger mechanism \cite{Sauter}. However, pair
creation in QED has never been observed directly although
planned new facilities such as a X-ray free electron laser (XFEL)
\cite{tesla,xfel} will allow
 to reach the region of required critical field strengths. Therefore
this non-perturbative effect was mainly studied
for applications providing strong enough fields
ranging from black hole quantum evaporation \cite{Damour} to particle
production in the early universe \cite{Parker} and in
ultra-relativistic heavy-ion collisions \cite{Casher}. 

An unsolved problem of conceptual and practical interest is the 
precise connection between field theory and kinetic theory. 
Recently a link between the mean
field approach of vacuum pair creation in a spatially homogeneous
Abelian 
background field 
and a kinetic formulation was established in \cite{kme}. The resulting
source term for
spontaneous pair creation is non-Markovian and 
retains quantum statistical effects  \cite{rau,sms}. In
many approaches the background field is treated as a time dependent
classical field with feedback incorporated via Maxwell's equation, e.g.
\cite{kkaj,jmeis,nayak,bloch,review}. In these approaches the
production of fermion/gluon pairs  was employed 
to describe the formation of a quark-gluon plasma. Herein we focus on the
production of bosonic particles in hot hadronic matter in QCD.

Lattice calculations, e.g. \cite{edwin}, as well as QCD Green function
approaches, e.g.
 \cite{review,alk3}, indicate that the
deconfinement and chiral phase transitions are coincident 
\cite{review,trento,njl,Bender}. 
At present it is an open question whether the restoration of the
U$_A$(1) symmetry which is broken in the QCD vacuum sets in already at
the deconfinement transition temperature or above.
 In addition   parity maybe 
spontaneously broken which is connected with a non-vanishing  QCD $\theta$ angle \cite{dkhar}.
The CP odd phase  is of particular interest since it may have experimental
signatures such as an enhanced production of $\eta$ and
$\eta'$ mesons \cite{jkap,alk1} which can contribute via their decays to
the low mass dilepton enhancement.  They can decay via  CP
violating  processes such as $ {\eta} \to {\pi}^0 {\pi}^0$.
%and of global parity odd asymmetries for charged pions.

Herein we study the production of $\eta'$-
particles during the decay of the CP-odd phase.
Complementary to \cite{dahr} where the decay rate of metastable
states was  estimated, we
study the full time evolution of the momentum distribution function
using a kinetic description based on an effective Lagrangian. We
start from  the  Witten-
DiVecchia-Veneziano model \cite{gvene}, however, different approaches
can be applied, e.g. \cite{alk}.

In this article, the external background field concept
is replaced by a  potential yielding self-interaction 
and non-linearity. This potential dominates the solution of the 
quantum kinetic equation which is derived using an evolution operator
approach. The introduced technique to link an effective
Lagrangian and kinetic theory is not restricted to the discussed
model calculation of  $\eta'$ production. It's application is general
in quantum field theory.

The article is organised as follows. In Section II we introduce the
model Lagrangian and identify the self-interaction parts.
In Section III we perform the quantization of the evolution operator
used in Section IV  to derive a quantum kinetic equation. In Sections
V and VI we discuss the decay of the CP odd phase in view of our
numerical results. 

%%%%%%%%%%%%%%%%%%%%%%%%%%%%%%%%%%%%%%%%%%%%%%%%%%%%%%%%%%%%%%%%%%%%%%%%%%%%%%
\section{The effective Lagrangian}
%%%%%%%%%%%%%%%%%%%%%%%%%%%%%%%%%%%%%%%%%%%%%%%%%%%%%%%%%%%%%%%%%%%%%%%%%%%%%%

We start from  the effective Lagrangian of the 
 Witten-DiVecchia-Veneziano model \cite{gvene}
\begin{eqnarray}
 {\cal L}_{eff} &=& \frac{f^2_{\pi}}{4}\bigg(tr(\partial_{\mu}U
        \partial_{\mu}U^+)  + tr(MU+MU^+) \nonumber \\ \label{lagr}
        & & -\frac{a}{N_c}[\theta - \frac{i}{2}tr(\ln U -\ln U^+)]^2\bigg),
\end{eqnarray}
 which describes the low-energy dynamics of the nonet of the 
 pseudoscalar mesons \cite{alk2} in the large $N_c$-limit of QCD.
The meson fields are described by 
 the $N_f \times N_f$-matrix $U$ in Eq.~(\ref{lagr}).
Explicit chiral symmetry breaking is realized  by the 
 current quark mass matrix $M$
 with the diagonal elements related to $\pi$ and $K$ meson masses.
 With the parametrization $U=\exp(i\phi/f_{\pi})$,
 the matrix $\phi$ representing the singlet and the octet meson fields
 yields the pseudoscalar nonet.
The last term in the effective Lagrangian  is related to  the
$U_A(1)$-anomaly: the singlet is massive also in chiral limit. 
The parameter $a=2 N_f \lambda_{YM}/f^2_{\pi}$
contains the topological susceptibility, $\lambda_{YM}$. Herein we focus on the
singlet state which is the main component for $\eta'$ and obtain the
following Lagrangian:
\begin{equation}
{\cal L}= \frac{1}{2} ({\partial}_{\mu} {\eta}) ({\partial}^{\mu}
{\eta}) + f^2 {\mu}^2 \cos\Big(\frac{\eta}{f}\Big) - \frac{a}{2} {\eta}^2\,.
\label{eq: odin}
\end{equation}
In Eq. (\ref{eq: odin}) $f=\sqrt{\frac{3}{2}} f_{\pi}$, where
$f_{\pi}=92$ MeV
is the semi-leptonic pion decay constant, ${\mu}^2= \frac{1}{3} (m^2_{\pi} +
2m^2_{K})$ is a parameter depending on $\pi$- and $K$-meson masses.
For zero temperature $T=0$, $a = m^2_{\eta} + m^2_{{\eta}^{\prime}}
-2 m_K^2 \simeq 0.726~ {\rm GeV}^2$ and 
${\mu}^2 \simeq 0.171~ {\rm GeV}^2$.
In response to non-zero temperature and density  mesons have an effective
mass, e.g. \cite{mesons}: $\mu$ and $a$ are functions of $T$ and hence 
the potential corresponding to (\ref{eq: odin}) has modified
properties close to the deconfinement phase transition. 

From (\ref{eq: odin}) we obtain the following Klein-Gordon type equation of motion for 
the field ${\eta}(\vec{x},t)$:
\begin{equation}\label{cur}
(\Box + m_0^2) {\eta} = J_s,
\end{equation}
where $m_0^2 \equiv a + {\mu}^2$. The nonlinear current
\begin{equation}
J_s \equiv -f{\mu}^2 \Big[ \sin\Big( \frac{\eta}{f} \Big) 
- \Big( \frac{\eta}{f}\Big) \Big]
\label{eq: pet}
\end{equation}
contains orders ${\eta}^3$ and higher and is
related to the self-interaction of the field $\eta$. Note
that the linear term of the total current $J = - {\mu}^2 {\eta}+J_s$ is
contained in the mass squared term of the lelf-hand side of Eq. (\ref{cur})

The total Hamiltonian density, ${\cal H} = {\cal H}_0 + {\cal H}_s$,
is given by 
\begin{eqnarray}\nonumber
{\cal H}_0 &=&\frac{1}{2} {\pi}^2 + \frac{1}{2}
(\vec{\nabla}{\eta})^2 + \frac{1}{2} m_0^2 {\eta}^2,\\
\label{eq: vosem}
{\cal H}_s &=&2f^2{\mu}^2 \Big[ {\sin}^2\Big(\frac{\eta}{2f}\Big) -
\Big(\frac{\eta}{2f}{\Big)}^2 \Big],
\end{eqnarray}
where ${\cal H}_0$ involves only the free field part with the mass
$m_0$; ${\cal H}_s$ includes self interaction starting at
orders $\eta^4$ and ${\pi}$ is the momentum canonically conjugate to $\eta$:  
\begin{equation}
{\pi}(\vec{x},t) = \dot{\eta}(\vec{x},t),
\label{eq: sem}
\end{equation}
where the overdot denotes the derivative with respect to time.
%%%%%%%%%%%%%%%%%%%%%%%%%%%%%%%%%%%%%%%%%%%%%%%%%%%%%%%%%%%%%%%%%%%%%%%%%%%%%%
\section{Evolution operator approach}
%%%%%%%%%%%%%%%%%%%%%%%%%%%%%%%%%%%%%%%%%%%%%%%%%%%%%%%%%%%%%%%%%%%%%%%%%%%%%%
We introduce the in-field, ${\eta}_{in}(\vec{x},t)$\footnote{The
model is defined in a finite volume: $V=L^3$, $-L/2 \leq x_i
\leq L/2$, $i=1,2,3$. The continuum limit is $\frac{1}{V}
\sum_{\vec{k}} \Longrightarrow \int \frac{d^3 \vec{k}}
{(2{\pi})^3}$. }, as a solution of Eq. (\ref{cur}) in absence
of sources and quantise it according to the standard 
canonical procedure (see Appendix A). The original 
self-interacting field is connected with the in-field by the 
unitary  transformation:
\begin{equation}
{\eta}(\vec{x},t) =U^{-1}(t) {\eta}_{in}(\vec{x},t) U(t),
\label{eq: odindevet}
\end{equation}
where
\begin{equation}
U(t) \equiv T\exp\{-i\int_{-\infty}^t dt^{\prime} H_s^{in}(t^{\prime})\}
\label{eq: dvanol}
\end{equation}
is the time evolution operator with
the self-interaction Hamiltonian written in terms of the
in-field operators 
\begin{equation}
H_s^{in} \equiv \int d^3x {\cal H}_s({\eta}={\eta}_{in} ;
{\pi}={\pi}_{in} )\,.
\label{dvaodin}
\end{equation}
In the limit $t \to -\infty$ we have $U(t) \to I$, so that
\begin{equation}
\lim_{t \to -\infty} {\eta}(\vec{x},t) = {\eta}_{in}(\vec{x},t).
\label{eq: dvatri}
\end{equation}
The exact meaning of (\ref{eq: dvatri}) 
depends on details of the current $J_s$ which in our model is
determined by  self-interaction taking place at {\it all} times.
Hence  Eq. (\ref{eq: dvatri}) is {\it a priori} difficult to justify.
%To proceed  we require that the limit will 
%be valid  for the matrix elements of the 
%field operators between normalized states. 
We assume 
an adiabatic vanishing of the interaction for  $t \to -\infty$.

The field ${\eta}(\vec{x},t)$ is given by the space-homogeneous mean value ${\phi}(t)=\langle {\eta}(\vec{x},t) \rangle$
and fluctuations ${\chi}$ 
\begin{equation}
{\eta}(\vec{x},t) = {\phi}(t) + {\chi}(\vec{x},t)
\label{eq: dvavosem}
\end{equation}
with $\langle {\chi}(\vec{x},t) \rangle =0$.
Assuming that ${\chi} \ll f$, quantum
fluctuations can be treated
perturbatively. Herein we  restrict ourselves to zeroth $^{(0)}$ and first $^{(1)}$
order. Substituting 
Eq. (\ref{eq: dvavosem})
into Eq. (\ref{cur}) yields
\begin{equation}
\label{box}
(\Box + m_0^2) {\chi} + \ddot{\phi} + m_0^2 {\phi} =J_s^{(1)},
\end{equation}
where
\begin{equation}
J_s^{(1)} \equiv J_s^{(0)} + {\mu}^2 \Big[ 1 - \cos\Big( \frac{\phi}{f} \Big)
\Big] {\chi}.
\label{eq: tridva}
\end{equation}
The zeroth order of the current  is given by
\begin{equation}
J_s^{(0)} \equiv -f{\mu}^2 \Big[ \sin\Big(\frac{\phi}{f} \Big) -
\Big( \frac{\phi}{f} \Big) \Big].    
\end{equation}
Taking the mean value $\langle
... \rangle$ of Eq. (\ref{box}), yields the vacuum mean field equation
\begin{equation}
\label{17}
\ddot{\phi} + a{\phi} + f{\mu}^2 \sin\Big( \frac{\phi}{f} \Big)=0.
\label{eq: tricet}
\end{equation}
Eq. (\ref{17}) in concert with Eq. (\ref{box}) provides the equation of motion for
the quantum fluctuations
\begin{equation}
(\Box + m_0^2) {\chi} = {{\mu}^2} \Big(1-\cos\Big(\frac{\phi}{f}\Big)
\Big) {\chi}.
\label{eq: tripet}
\end{equation}
The right-hand side of this equation vanishes in the in-limit.
Rewriting (\ref{eq: tripet}) for the Fourier components ${\chi}(\vec{k},t)$,
we obtain a Mathieu type equation  \cite{land,tras}
\begin{equation}
\ddot{\chi}(\vec{k},t) + {\omega}_k^2(t) {\chi}(\vec{k},t)=0,
\label{eq: trishest}
\end{equation}
where 
\begin{equation}
{\omega}_k^2(t) \equiv ({\omega}_k^0)^2 - {\mu}^2
\Big( 1 - \cos\Big(\frac{\phi}{f}\Big) \Big)
\label{eq: trisem}
\end{equation}
and ${\omega}_k(t)$ is the time-dependent frequency of the
fluctuations with $\lim_{t \to -\infty} {\omega}_k(t) = {\omega}_k^0=\sqrt{k^2+m_0^2}$.

For $a>{{\mu}^2}$, the frequency squared
is positive for all momentum modes and at all times. However, if
$a<{{\mu}^2}$, 
${\omega}_k^2(t)$ can be negative  for  modes
below a critical momentum $\vec{k}_c$ indicating a tachyonic regime.

It is important to observe that Eqs. (\ref{eq: tricet}) and 
(\ref{eq: tripet}) are coupled.
Although the fluctuations do not react on the vacuum mean
field, the latter modifies the equation for fluctuations
via a time dependent frequency.

The self-interaction Hamiltonian density corresponding to the
equations (\ref{eq: tricet}) and (\ref{eq: tripet}) is quadratic in ${\chi}$,
\begin{eqnarray}\nonumber
{\cal H}_s^{(1)} = &&2f^2{\mu}^2 \big[ {\sin}^2\Big( \frac{\phi}{2f} \Big) -
\Big( \frac{\phi}{2f}{\Big)}^2 \Big]\\\nonumber
& +& 
f{\mu}^2 \Big[ \sin\Big( \frac{\phi}{f} \Big)
- \Big( \frac{\phi}{f}\Big) \Big] {\chi}\\
&+& \frac{1}{2} {\mu}^2 \Big[ \cos\Big( \frac{\phi}{f} \Big) -1 \Big]
{\chi}^2,
\label{ham}
\end{eqnarray}
and also vanishes when $t \to -\infty$. Hence in the approximation of
preserving quantum fluctuations in the vacuum mean field but neglecting 
the feedback,
the adiabatic
hypothesis of vanishing interactions for $t \to -\infty$ discussed
with Eq. (\ref{eq: dvatri}) is justified.

For the Fourier components of the fluctuations, we write the
ansatz analogous to (\ref{eq: odinodin}),
\begin{equation}
{\chi}(\vec{k},t) = {\Gamma}_{\vec{k}}(t) a(\vec{k},t) +
{\Gamma}_{\vec{k}}^{\star}(t) a^{\dagger}(-\vec{k},t),
\label{eq: cetodin}
\end{equation}
where
\begin{equation}
{\Gamma}_{\vec{k}}(t) = \frac{1}{\sqrt{2{\omega}_k(t)}}
\exp\{ -i{\Theta}_k({\omega}_k,t) \},
\label{eq: cetdva}
\end{equation}
and ${\Theta}_k({\omega}_k,t)$ is a phase which in the in-limit
takes the form ${\omega}_k^0 t$. In the same
limit, ${\Gamma}_{\vec{k}}(t) \to {\Gamma}_{\vec{k}}^0(t)$,
while the time-dependent operators $a(\vec{k},t)$,
$a^{\dagger}(\vec{k},t)$ with $\lim_{t \to -\infty} a(\vec{k},t) =
a_{in}(\vec{k})$ and $\lim_{t \to -\infty} a^{\dagger}(\vec{k},t) = 
a^{\dagger}_{in}(\vec{k})$.

In the case when the fluctuations and the frequency ${\omega}_k$
vary adiabatically slowly in time, the dynamical phase ${\Theta}_k$
can be chosen as
\begin{equation}
{\Theta}_k^{ad} = \int^t {\omega}_k(t^{\prime}) dt^{\prime}.
\label{eq: cetpet}
\end{equation}
The relations between the Fourier components ${\eta}(\vec{k},t)$
and ${\chi}(\vec{k},t)$  and the corresponding conjugate momenta are
given by
\begin{eqnarray}
{\eta}(\vec{k},t)& =& 
{\chi}(\vec{k},t) 
+ {\delta}_{\vec{k} ,0} \sqrt{V} {\phi}(t)\,,
\label{eq: cetshest}\\
{\pi}(\vec{k},t)& =& {\pi}_{\chi}(\vec{k},t)
+ {\delta}_{\vec{k},0} \sqrt{V} \dot{\phi}(t)\,.
\label{eq: cetsem}
\end{eqnarray}
The Fourier components of the operator
${\pi}_{\chi}$ are
\begin{equation}
{\pi}_{\chi}(\vec{k},t) = -i{\omega}_k(t) \Big[ {\Gamma}_{\vec{k}}(t)
a(-\vec{k},t) - {\Gamma}_{\vec{k}}^{\star}(t)
a^{\dagger}(\vec{k},t) \Big]
\label{eq: cetvosem}
\end{equation}
and in the limit $t \to -\infty$ this ansatz reduces to 
(\ref{eq: odinshest}).

Using  Eqs. (\ref{eq: cetodin}) and (\ref{eq: cetvosem}), we obtain 
the following
relations between $a(\vec{k},t)$, $a^{\dagger}(\vec{k},t)$ and the in-operators:
\begin{eqnarray}\nonumber
a(\vec{k},t)& =& \frac{1}{2{\Gamma}_{\vec{k}}(t)} \{
U^{-1}(t) \Big[ {\eta}_{in}(\vec{k},t) + \frac{i}{{\omega}_k}
{\pi}_{in}(-\vec{k},t) \Big] U(t))\\
&&- {\delta}_{\vec{k},0} \sqrt{V} \Big( {\phi} + \frac{i}{{\omega}_0}
\dot{\phi} \Big) \},
\label{eq: petodin}\\\nonumber
a^{\dagger}(\vec{k},t)& = &\frac{1}{2{\Gamma}_{\vec{k}}
^{\star}(t)} \{ U^{-1}(t) \Big[ {\eta}_{in}(-\vec{k},t) -
\frac{i}{{\omega}_k} {\pi}_{in}(\vec{k},t) \Big] U(t)\\
&-& {\delta}_{\vec{k},0} \sqrt{V} \Big( {\phi} - \frac{i}{{\omega}_0}
\dot{\phi} \Big) \},
\label{eq: petdva}
\end{eqnarray}
where ${\omega}_0 \equiv {\omega}_{k=0}$. It is easy  to verify
that the operators $a(\vec{k},t)$,
$a^{\dagger}(\vec{k},t)$ 
fulfill the same commutation relations as the in-operators, Eqs. (\ref{eq: odindva}).
Hence the transformation defined by the evolution operator is
canonical.
The ${\phi}$-dependent terms in Eqs. (\ref{eq: petodin})-
(\ref{eq: petdva}) act like counter terms
which cancel the vacuum mean field contribution of the previous
terms, so that  (\ref{eq: petodin}) and 
(\ref{eq: petdva}) do not depend on ${\phi}$ explicitly.

%%%%%%%%%%%%%%%%%%%%%%%%%%%%%%%%%%%%%%%%%%%%%%%%%%%%%%%%%%%%%%%%%%%%%%%%%%%%%%
\section{Kinetic equation}
%%%%%%%%%%%%%%%%%%%%%%%%%%%%%%%%%%%%%%%%%%%%%%%%%%%%%%%%%%%%%%%%%%%%%%%%%%%%%%
The number of particles of a given state characterized by the momentum
$\vec{k}$ at
time t is given by
\begin{equation}
{\cal N}(\vec{k},t) \equiv \langle 0|a^{\dagger}
(\vec{k},t) a(\vec{k},t) |0 \rangle .
\label{eq: pettri}
\end{equation}
In the limit $t \to -\infty$, ${\cal N}(\vec{k},t)$ tends of course  towards
the occupation number density of the in-field:
\begin{equation}
{\cal N}(\vec{k},t) \to N(\vec{k}) \equiv \langle 0| a_{in}^{\dagger}(\vec{k})
a_{in}(\vec{k})|0 \rangle.
\label{eq: petcet}
\end{equation}
Substituting (\ref{eq: petodin})
and (\ref{eq: petdva}) into (\ref{eq: pettri}) 
and introducing the instantaneous states
$U|0 \rangle \equiv |U \rangle$, $\langle 0| U^{-1} \equiv \langle U|$,
the particle number can be written as
\begin{eqnarray}\nonumber
{\cal N}(\vec{k},t)&=& \frac{{\omega}_k}{2} \langle U|
{\eta}_{in}^{\dagger}(\vec{k},t) {\eta}_{in}(\vec{k},t) 
+ \frac{1}{{\omega}_k^2}  {\pi}_{in}(\vec{k},t)
{\pi}_{in}^{\dagger}(\vec{k},t) |U \rangle\\\nonumber
&+& \frac{i}{2} \langle U| {\eta}_{in}^{\dagger}(\vec{k},t)
{\pi}_{in}^{\dagger}(\vec{k},t) - {\pi}_{in}(\vec{k},t)
{\eta}_{in}(\vec{k},t) |U \rangle\\
&-& {\delta}_{\vec{k},0} \frac{{\omega}_0}{2} V
( {\phi}^2 + \frac{1}{{\omega}_0^2} \dot{\phi}^2).
\label{eq: petshest}
\end{eqnarray}
The number of particles of momentum
$\vec{k}$ is not equal to that of momentum $(- \vec{k})$ for all
times $t$.
Therefore it is convenient to introduce
\begin{equation}
{\cal N}_{\pm}(\vec{k},t) \equiv \frac{1}{2}
\Big( {\cal N}(\vec{k},t) \pm {\cal N}(-\vec{k},t) \Big),
\label{eq: petsem}
\end{equation}
where ${\cal N}_{+}(\vec{k},t)$ is the particle number
averaged over the directions $\vec{k}$ and $(-\vec{k})$,
while ${\cal N}_{-}(\vec{k},t)$ measures the degree of  asymmetry.
At fixed volume, the occupation number densities can change in time for two
reasons, either with the change of the number of particles
or with the change of the vacuum state. The presence of the
background field leads to a restructuring of the vacuum
state. Note that in the case when the background is a constant classical 
field, the definition of the vacuum does not change in time.
One considers excitations with respect to this vacuum and
interprets an increase in the occupation number density as 
particle production. The vacuum state itself is ``empty'',
i.e. without particles.

In our model, the background field ${\phi}(t)$ is 
 periodic in time, i.e. in addition to the quantum fluctuations around
${\phi}$ we have oscillations of ${\phi}$ itself. Therefore
 the vacuum restructures itself 
at each moment in time and consequently the occupation number
density has to be redefined as well since it is assumed to be zero
only for the vacuum state.
In the time
evolution of the densities ${\cal N}_{\pm}(\vec{k},t)$,
it is therefore necessary to separate the contribution of the real
particle production from the one related to the vacuum state
redefinition. This is achieved by using the expansion 
(\ref{ham}) in the evolution operator $U(t)$.

We consider first the time evolution of ${\cal N}_{-}
(\vec{k},t)$. Taking the time derivative of ${\cal N}_{-}(\vec{k},t)$
and taking into account the relation $i\dot{U} = H_s^{in} U$
we find:
\begin{eqnarray}\label{eq: shestodin}
&&\dot{\cal N}_{-}(\vec{k},t) =\\
&& -\frac{1}{2} \langle U|
\Big[ H_s^{in} , {\eta}_{in}^{\dagger}(\vec{k},t)
{\pi}_{in}^{\dagger}(\vec{k},t) - {\pi}_{in}(\vec{k},t)
{\eta}_{in}(\vec{k},t) {\Big]}_{-} |U \rangle .
\nonumber
\end{eqnarray}
Using the expansion (\ref{ham}),
the commutator in (\ref{eq: shestodin}) is readily calculated:
\begin{equation}
\dot{\cal N}_{-}(\vec{k},t) = \frac{1}{\sqrt{V}} {\rm Im}
\int d^3x \langle 0| e^{i\vec{k}\vec{x}} {\chi}(\vec{k},t)
J_s^{(1)}(\vec{x},t) |0 \rangle.
\label{eq: shestdva}
\end{equation}
The time evolution of the density ${\cal N}_{-}
(\vec{k},t)$ is determined by the self-interaction of the field
${\eta}(\vec{x},t)$. To get an exact formula valid in all orders
of perturbations in $({\chi}/f)$ it is sufficient to replace
$J_s^{(1)}$ by $J_s$ in Eq. (\ref{eq: shestdva}).

In the chosen approximation of small quantum fluctuations, the current
$J_s$ is considered in  first order in ${\chi}$. 
Using Eq. (\ref{eq: tridva}), the integral in Eq. (\ref{eq: shestdva})
turns out to be real and one obtains
$\dot{\cal N}_{-}(\vec{k},t)=0$. In first order in
${\chi}$ the number density ${\cal N}_{-}(\vec{k},t)$ is therefore
conserved. 

Taking the time derivative of
${\cal N}_{+}(\vec{k},t)$ we obtain the evolution equation
\begin{eqnarray}\nonumber
&&\dot{\cal N}_{+}(\vec{k},t) =\frac{\dot{\omega}_k}{{\omega}_k}
{\rm Re} \Big[ C(\vec{k},t) e^{-2i{\Theta}_k} \Big]\\\nonumber
&+&\frac{1}{{\omega}_k} ({\omega}_k^2 - ({\omega}_k^0)^2)
{\rm Im} \Big[ C(\vec{k},t)e^{-2i{\Theta}_k} \Big]-\frac{1}{{\omega}_0} {\delta}_{\vec{k},0} V J_s^{(0)} \dot{\phi}\\
&+&  \frac{i}{2{\omega}_k} \langle U| \Big[ H_s^{in},
{\pi}_{in}(\vec{k},t) {\pi}_{in}^{\dagger}(\vec{k},t)
{\Big]}_{-} |U \rangle,
\label{eq: shesttri}
\end{eqnarray}
where we have defined the time-dependent pair correlation
function $C(\vec{k},t) \equiv \langle 0| a(-\vec{k},t)a(\vec{k},t) |0 \rangle.$
Calculating the commutator in the expression for
$\dot{\cal N}_{+}(\vec{k},t)$, we find
\begin{eqnarray}\nonumber
&&\frac{i}{2{\omega}_k} \langle U| \Big[ H_s^{in},
{\pi}_{in}(\vec{k},t) {\pi}_{in}^{\dagger}(\vec{k},t)
{\Big]}_{-} |U \rangle= \\\nonumber
&&\frac{1}{\sqrt{V}} \frac{1}{{\omega}_k} {\rm Re}
\int d^3x \langle 0| e^{i{\vec k}{\vec x}} {\pi}_{\chi}^{\dagger}
(\vec{k},t) J_s^{(1)}(\vec{x},t) |0 \rangle\\
& +& \frac{1}{{\omega}_0} {\delta}_{\vec{k},0} V J_s^{(0)} \dot{\phi}.
\label{eq: shestshest}
\end{eqnarray}
With Eq. (\ref{eq: tridva}), this last expression becomes
\begin{eqnarray}\nonumber
&&
\frac{i}{2{\omega}_k} \langle U| \Big[ H_s^{in} ,
{\pi}_{in}(\vec{k},t) {\pi}_{in}^{\dagger}(\vec{k},t)
{\Big ]}_{-} |U \rangle  = \\\nonumber
&&\frac{{\mu}^2}{2{\omega}_k} \Big( 1-\cos\Big( \frac{\phi}{f} \Big) \Big)
{\rm Im}\Big[ C(\vec{k},t)e^{-2i{\Theta}_k} \Big]\\
&+& \frac{1}{{\omega}_0} {\delta}_{\vec{k},0} V J_s^{(0)} \dot{\phi},
\label{eq: shestsem}
\end{eqnarray}
so that in the small quantum fluctuations approximation
\begin{equation}
\dot{\cal N}_{+}(\vec{k},t) = \frac{\dot{\omega}_k}{{\omega}_k}
{\rm Re} \Big[ C(\vec{k},t)e^{-2i{\Theta}_k} \Big] .
\label{eq: shestvosem}
\end{equation}
In the same approximation, the pair correlation function
$C(\vec{k},t)$ obeys the equation
\begin{eqnarray}\nonumber
&&\dot{C}(\vec{k},t) - 2i (\dot{\Theta}_k - {\omega}_k) C(\vec{k},t)
=\\
&& \frac{\dot{\omega}_k}{2{\omega}_k} \Big( 1+2{\cal N}_{+}
(\vec{k},t) \Big)  e^{2i{\Theta}_k}.
\label{eq: shestdevet}
\end{eqnarray}
Its formal solution is
\begin{eqnarray}\nonumber
C(\vec{k},t) =&& e^{2i{\Theta}_k} \int_{-\infty}^{t} dt^{\prime}
 \frac{\dot{\omega}_k(t')}{2{\omega}_k(t')}
\Big( 1+ 2{\cal N}_{+}(\vec{k},t^{\prime}) \Big)\times \\
&&e^{2i({\Theta}^{ad}_k(t^{\prime}) - {\Theta}^{ad}_k(t) )}.
\label{eq: semnol}
\end{eqnarray}
Substituting it into Eq. (\ref{eq: shestvosem}), we obtain a 
closed equation for ${\cal N}_{+}(\vec{k},t)$ similar to \cite{kme}:
\begin{eqnarray}\nonumber
\dot{\cal N}_{+}(\vec{k},t) = &&
\frac{\dot{\omega}_k}{2{\omega}_k}
\int_{-\infty}^{t} dt^{\prime}  \frac{\dot{\omega}_k(t')}
{{\omega}_k(t')} \Big( 1 + 2{\cal N}_{+}(\vec{k},
t^{\prime}) \Big) \times\\
&& \cos[2{\Theta}_k^{ad}(t) - 2{\Theta}_k^{ad}(t^{\prime})].
\label{eq: semodin}
\end{eqnarray}
Eq. (\ref{eq: semodin}) is a quantum kinetic equation which determines
the time evolution of the number of particles of a fixed
momentum $\vec{k}^2 >\vec{k}^2_c$.
Note that the background field does not contribute to the
kinetic equation directly, but only via the frequency of the
quantum fluctuations. Therefore, the change of 
${\cal N}_{+}(\vec{k},t)$ in time in Eq. (\ref{eq: semodin})
is due to particle production during the
fluctuations.

In the regime of the negative frequency squared, when
$\vec{k}^2 < \vec{k}^2_c$ and ${\omega}_k = \pm i {\nu}_k
\equiv \pm i \sqrt{\vec{k}^2_c - \vec{k}^2}$, one of the
phase factors in the ansatz (\ref{eq: cetodin}), ${\Gamma}_{\vec{k}}(t)$
or ${\Gamma}_{\vec{k}}^{\star}(t)$, grows exponentially in time.
Instead of oscillations we have an exponential growth of long
wavelength quantum fluctuations with momenta $\vec{k}^2 <
\vec{k}^2_c$. This is the so-called tachyonic instability \cite{kirz,anse,linde}.

\begin{figure}[t]
\centerline{\epsfig{figure=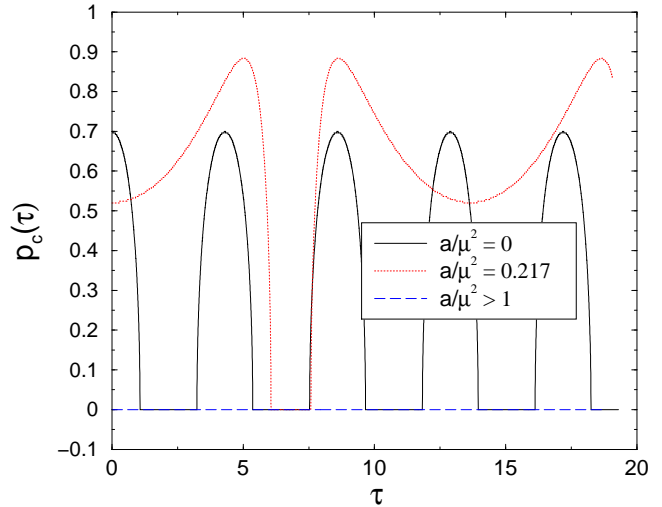,width=7.0cm,angle=-90}}
\caption{The dependence of the critical momentum $p_c=k_c/\mu$,
  Eq. {\protect (\ref{kc})}, on time $\tau=t\,\mu$. A
  non-vanishing value indicates the appearance of tachyonic modes. The
  time dependence of $p_c$ is due to the alternating $\phi$-field and
  depends strongly on the choice of the initial values.}
\label{fig4}
\end{figure}

Such a tachyonic regime is realized for potential paramters 
 $a/{{\mu}^2}<1$.
Whether the system evolves in the tachyonic or non-tachyonic regime 
is dynamically fixed by the time dependent critical momentum:
\begin{eqnarray}
\label{kc}
\vec{k}^2_c(t) & = &\left\{
\begin{array}{lcl}
{{\mu}^2} |\cos({\phi}/f)| -a\,, && {{\mu}^2}\cos({\phi}/f) + a<0\\
0\,, &&  {\rm otherwise}
\end{array}\right.
\end{eqnarray}
plotted in Fig. \ref{fig4} for different parameters $a/{{\mu}^2}$. 
For $a/{{\mu}^2} > 1$ the critical momentum is zero since the
frequency is always positive and no tachyonic modes can establish. 
The critical momentum for $a/{{\mu}^2}<1$ oscillates
in tune with the time dependence of the vacuum mean field
$\phi$. The time evolution shows that the same momentum state can
change its nature during the evolution. 
 In that case a different kinetic equation must be derived and solved
 which evolves all tachyonic modes in time.  
Therefore the analytical and numerical treatment is a complicated
challenge and a
quantitative analysis of  $a/{{\mu}^2}<1$ states will be provided elsewhere.

The total number of particles of all modes at any time
$t$ is given by
\begin{equation}
{\cal N}(t) = 2\int_0^\infty\frac{d^3{\vec k}}{(2\pi)^3} {\cal N}_{+}(\vec{k},t).
\label{eq: semcet}
\end{equation}
Simple power counting yields that this expression is finite.

%%%%%%%%%%%%%%%%%%%%%%%%%%%%%%%%%%%%%%%%%%%%%%%%%%%%%%%%%%%%%%%%%%%%%%%%%%%%%%
\section{Decay of the CP odd phase}
%%%%%%%%%%%%%%%%%%%%%%%%%%%%%%%%%%%%%%%%%%%%%%%%%%%%%%%%%%%%%%%%%%%%%%%%%%%%%%
In the vicinity of the phase transition the QCD vacuum rearranges: chiral
symmetry breaking and confinement drive quark matter into hadronic
states.
In this region topological phenomena such as the appearance of a
non-vanishing QCD $\theta$ angle may occur.
As a consequence the CP symmetry is dynamically broken and in these 
regions a CP-odd phase can occur (paramater Set IV in Table I). In the rapid cooling of the 
hot QCD matter down to the critical temperature $T_d$, the U$_A$(1) 
breaking gets restored either completely (Set I) or partially (Sets II, 
III).
After such a quench of the effective potential (\ref{eq: sempet}), the system is in 
the false vacuum state.
The decay of this CP odd phase is a time dependent process and 
can be study within the kinetic approach introduced in the previous
section.

\begin{figure}[t]
\centerline{\epsfig{figure=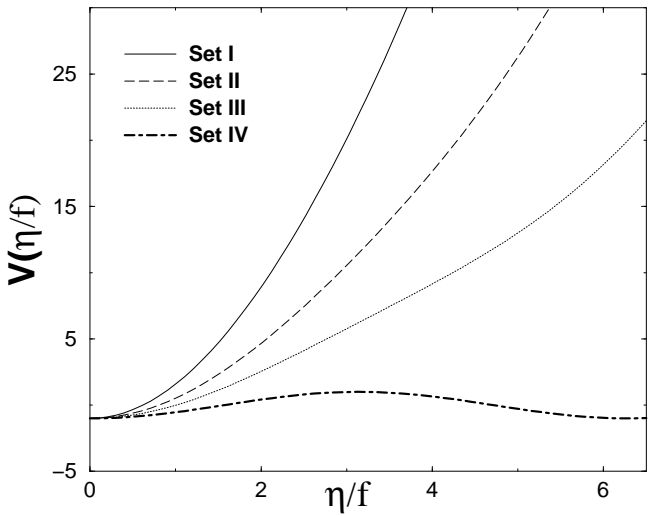,width=7.0cm,angle=-90}}
\caption{The shape of the potential $V({\eta}/f)$, {\protect (\ref{eq: sempet})}, is plotted 
 for different
values of $a/{{\mu}^2}$. Local minima characteristic for $a/{{\mu}^2}<1$ assumed at large 
temperatures disappear due to an applied fast quench. 
\label{fig1}}
\end{figure}

The potential of the effective Lagrangian density (\ref{eq: odin})
\begin{equation}
V(\eta/f) \equiv - \cos\Big( \frac{\eta}{f} \Big) +
\frac{a}{2\mu^2} (\eta/f)^2
\label{eq: sempet}
\end{equation}
is plotted in Fig. \ref{fig1} for different values of the potential
parameter, $a/{{\mu}^2}$, according to Table \ref{table}.

\begin{table}
\begin{center}
\begin{tabular}{c|c|c|c|c}
Set &I&II&III&IV\\\hline
$a/\mu^2\bigg/(a/\mu^2)_{vac}$ &1& 1/2 & 1/4&0
\end{tabular}
\caption{Different values of $a/\mu^2$ as used in the numerical
  calculation.  ($a/\mu^2)_{vac}\sim 4.24$ is the vacuum value for
which all mesons have their vacuum masses. 
Set I assumes a fast quench after which the vacuum value is 
immediately reached, i.e. the $\eta'$ mass assumes its vacuum value in
the vicinity of $T_d$.
 This scenario is compared with slow quenches corresponding to
 parameters given in Set II and Set III. Set IV leads to the
 appearance of tachyonic modes; a value only possible for $T>T_d$.
\label{table}} 
\end{center}
\end{table}
Starting from the  quark gluon plasma phase  in which  $a/{{\mu}^2}$
is suppressed, Set IV in Table \ref{table}, the potential 
changes from the cosine shape to a parabolic shape due a sudden quench
at the deconfinement phase transition. The metastable states located
in the local minima of the potential roll
smoothly back into the trivial minimum and oscillate around it. This
situation is 
formalized in  assumption (\ref{eq: dvavosem}): 
The ${\eta}$-field
can be  decomposed into its vacuum mean value
${\phi}(t)$ and its quantum fluctuations ${\chi}$.
During  the decay, energy is transferred
from ${\phi}$ to ${\chi}$. As a result, ${\phi}$
is damped, while the number of particles in quantum
fluctuations increases. It is assumed that during this
process the temperature does not change essentially,
and  particle production proceeds in a fixed
potential characterized by $a/{{\mu}^2}$. This
process takes place on  a time scale typical for the hadronisation
process: $1 - 10$ fm/c.
\begin{figure}[t]
\centerline{\epsfig{figure=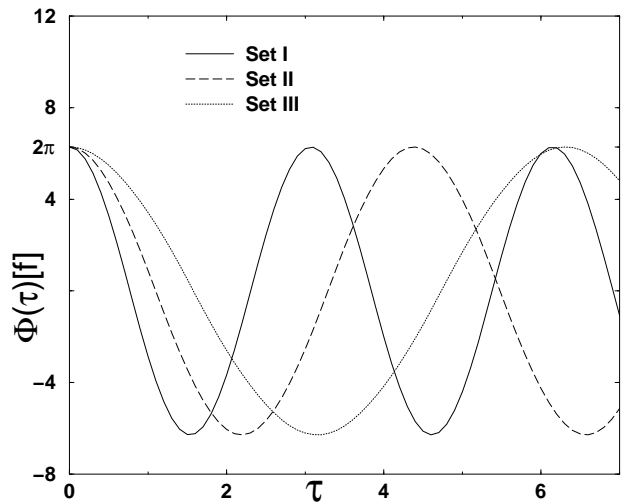,width=7.0cm,angle=-90}}
\caption{The solution of the vacuum mean field equation as function of
  time, Eq. {\protect
    (\ref{eq: semsem})}, is shown 
for different values of $a/{{\mu}^2}$ ({\it c.f.} Table \ref{table}) for the initial conditions
$\phi(0)/f=2\pi$ and ${\dot \phi}(0)/f=0$. Note that for
$a/\mu^2=0$, $\phi(\tau)$ would be  constant.}
\label{fig2}
\end{figure}
The  potential 
parameter $a/\mu^2$ depends on
temperature due to medium dependent meson masses. However its exact behaviour near the critical
temperature 
is unknown. Lattice calculations as well as QCD models suggest that $\pi$, $K$
and $\eta$ meson properties  have only a weak dependence on $T$. About the
response of $\eta'$ to increasing $T$ is much less  known and therefore we explore different 
scenarios summarised in Table \ref{table}. Set I assumes that the
medium dependence is negligible.  Set II (III, IV) corresponds to an in-medium
reduction  of the $\eta'$ mass of about 20 (40, 60)$\%$ applying the
simple equations given in connection with Eq. (\ref{eq: odin}),
\cite{alk1}.  
It is important to note that $f\sim f_\pi$ can be considered as an order
parameter for the chiral phase transition and hence is strongly
suppressed at $T_d$, i.e $f(T\sim T_d)=0.1\,f(T=0)$. In the herein
applied scenario, 
the in-medium dependent parameters $a/\mu^2$ and $f$  change only during the 
fast quench. Their time-dependence can therefore be neglected. 

The solution of the non-linear Eq. (\ref{eq: semsem}) for ${\phi}({\tau})$ 
with different  $a/{{\mu}^2}$ is  plotted in
Fig. \ref{fig2}, employing the initial conditions 
${\phi}(0)/f = 2\pi$ and  $(d{\phi}(\tau)/d{\tau})_{{\tau}
=0} =0$ throughout the numerical calculations.
We see that the period and the amplitude
of the oscillations of ${\phi}({\tau})$ vary with the change 
of $a/{{\mu}^2}$.
The oscillations are not damped since
feedback of the fluctuations on the mean field is  neglected. 
The field $\phi$ provides the background
field for the solution of  the quantum kinetic equation given in 
Eq. (\ref{eq: semdevet}).

%%%%%%%%%%%%%%%%%%%%%%%%%%%%%%%%%%%%%%%%%%%%%%%%%%%%%%%%%%%%%%%%%%
\section{Numerical results}
%%%%%%%%%%%%%%%%%%%%%%%%%%%%%%%%%%%%%%%%%%%%%%%%%%%%%%%%%%%%%%%%%%
The solution of the quantum kinetic equation (\ref{eq: semodin}) describes the
production of $\eta'$ particles: the  momentum dependence and it's time
evolution. The strong back ground field, $\phi$, leading to a sizeable 
particle production rate is given by the solution of the non-linear
Eq. (\ref{eq: tricet}), see Fig. \ref{fig2}.
\begin{figure}[t]
\centerline{\epsfig{figure=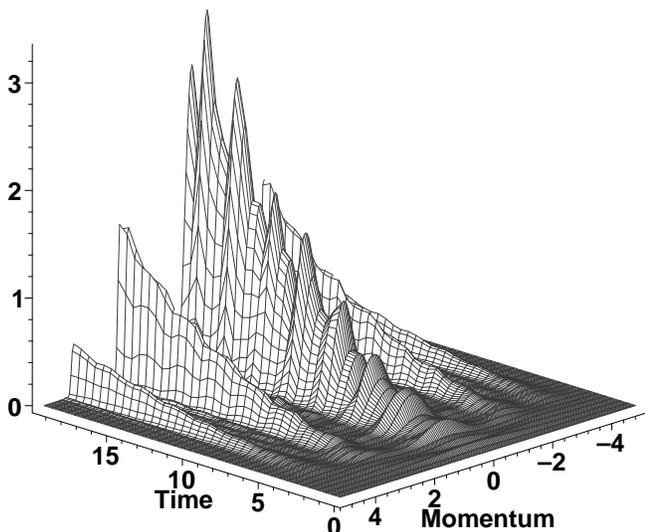,width=7.0cm,angle=-90}}
\caption{The time evolution of the momentum distribution function for
  parameter Set III. Most of the mesons are produced with small momenta but additional resonance bands appear for larger momenta; their maximal amplitude
is smaller. The time evolution is characterized by an increase of the
particle number and a repeated spike structure. }
\label{fig3d}
\end{figure}

We perform the numerical calculation  using dimensionless variables
and solve the kinetic equation as a system of
coupled differential equations, Eqs. (\ref{eq: vosemshest}-\ref{eq:
  vosemvosem}),  introduced in Appendix B.
The decay starts at $\tau=0$, for which ${\cal N}_+(\vec{p},0)=0; 
\phi(0)=2 \pi f$ and ${\dot \phi} (0)=0$ define the initial conditions.

As result we obtain
the number of particles produced during the decay of the CP odd phase
in the false vacuum. Herein we restricted ourselves to the study of 
 the non-tachyonic regime, i.e. we
explore the solution for positive frequencies, Eq. (\ref{eq:
  vosemodin}),  
corresponding to $a/{{\mu}^2}>1$ given in Table \ref{table}. 
\begin{figure}[t]
\centerline{\epsfig{figure=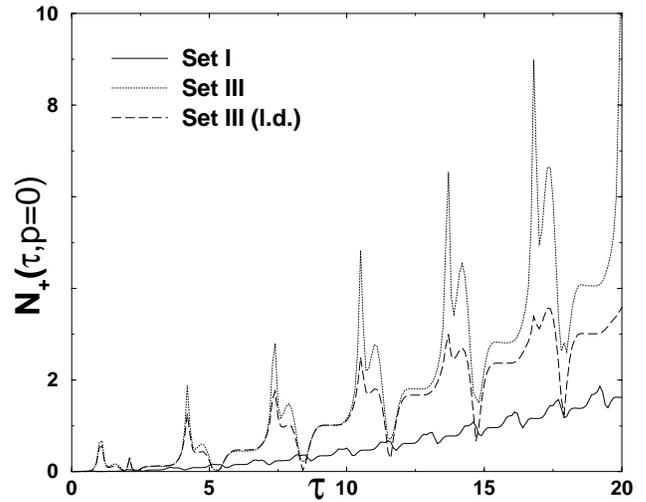,width=7.0cm,angle=-90}}
\caption{The time evolution of the particle number for two different
  $a/{{\mu}^2}>1$  when the system is in the
  non-tachyonic regime, {\protect (\ref{eq: semdevet})}. The double
  spike structure on top of the rapid growth repeats periodically in
  tune with the mean field's frequency ({\it c.f.} Fig. {\protect \ref{fig2}}). An estimate in low density approximation shows that
  inclusion of the Bose quantum statistics leads to a pronounced enhancement.}
\label{fig3}
\end{figure}
In Fig. \ref{fig3d} we show the complete numerical solution. Two
features are apparent: 
(i) the fast increase is characterised by a repeated structure on top
of the curve, 
(ii) additional to the occupation of low momentum states we observe
the appearance of resonance bands at larger momenta.

In Fig. \ref{fig3} we
plot the time evolution of the particle number for zero momentum and
compare the solution for Set I and Set III. We
observe a very fast increase of the number of produced particles.
A maximum occupation of a given momentum state at a given time is
reached for small values of the potential parameter, i.e. using Set
III. 
For larger values of the
potential parameter, e.g. Set I, less particles are produced in a given time since the source term 
is suppressed by a larger mass term $a/\mu^2$ in $\omega(p)$, see (\ref{eq: vosemodin}).

The most striking feature in this plot is
the periodically repeated spike structure on top of the overall
growth, ({\it c.f.} \cite{felder}). This pattern appears with the same
frequency as the background
field oscillates, see Fig. \ref{fig2}. When back reactions are
included this would possibly not be the case. The spike structure is
smoother for Set I compared to Set III but still
characteristic for the evolution. 

Herein we also compare the full
solution with the low density approximation (l.d.). The low density
approximation assumes that ${\cal N}_+({\vec p},t) \ll 1$ and hence suggests that the
solution of the kinetic equation does not depend on the pre-history of
the systems evolution. Any calculation which does not retain quantum
statistical 
effects necessarily employs this ansatz. From Fig. \ref{fig3} it is
plain that the inclusion of the Bose statistical factor  into the 
kinetic equation leads to Bose enhancement as soon as $N_+
\sim 1$, appearing at $\tau \sim 1$. This effect becomes more
pronounced with increasing time. 
\begin{figure}[t]
\centerline{\epsfig{figure=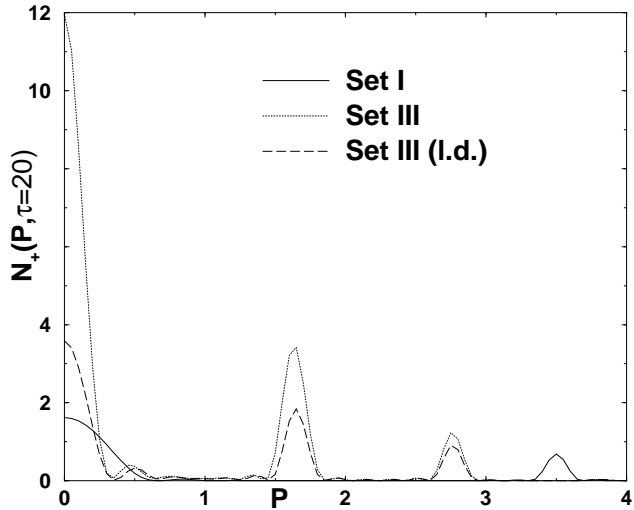,width=7.0cm,angle=-90}}
\caption{The particle number as function of momentum, $p=|{\vec p}|$, for
  two different $a/{{\mu}^2}>1$.  Bose enhancement of mainly the low
  momentum states is apparent. A characteristic second resonance band
  appears for large momenta.}
\label{fig6}
\end{figure}
The momentum dependence at a given time, Fig. \ref{fig6}, shows that most of the
particles are produced with small momenta. Additional resonance bands
appear. The smaller the value of the potential parameter is reached in
the quench the closer
the second maxima appears to the first one.
The reason for this resonance effect is typical for the Mathieu type
equation: the two intrinsic
frequencies of the background field and of the production process 
are of the same order of magnitude and resonances are likely to appear.

In Fig. \ref{fig6} we also compare the
momentum dependence of the full non-Markovian solution with a
calculation in low density limit. It is apparent that the Bose
enhancement acts naturally on the lower momenta. For the considered
case the occupation number is enhanced by a factor of 4. For large momenta details of
the quantum statistics are suppressed. Therefore the higher resonance
bands  are much less  affected by
quantum corrections. 
It is plain from this study that quantum statistical effects
cannot be neglected: the low density approximation is invalid if the
produced number density exceeds a critical value at very early times
of the evolution.

%%%%%%%%%%%%%%%%%%%%%%%%%%%%%%%%%%%%%%%%%%%%%%%%%%%%%%%%%%%%%%%%%%%%%%%%%%%%%%
\section{Summary}
%%%%%%%%%%%%%%%%%%%%%%%%%%%%%%%%%%%%%%%%%%%%%%%%%%%%%%%%%%%%%%%%%%%%%%%%%%%%%%
Starting from the singlet  Witten-DiVecchia-Veneziano effective
Lagrangian we have derived a quantum kinetic equation describing the
production of  ${\eta}'$-mesons from a CP-odd metastable vacuum state. We
have employed a general method based on the evolution operator holding
also for other model langrangians.
The vacuum mean field provides a classical, self-interacting strong
background field. Quantum fluctuations around the dynamical
mean field value are considered but their  feedback to the background
field is neglected. Due to these quantum fluctuations particles are
produced and the time evolution of this process is described by a
non-Markovian equation for the distribution function of the produced
$\eta'$ mesons.

We find that the details of the decay process depend strongly on the
applied quench. The number of produced particles is much larger when
the $\eta'$ mass is suppressed in the vicinity of the phase boundary.
Most of the
particles are produced  with low momenta, for large momenta additional
resonances appear. Furthermore, we have demonstrated that quantum statistical effects are
important and lead to a pronounced enhancement of the particle
occupation number for low momenta. In the case $a/\mu^2<1$, tachyonic instabilities occur for momenta 
smaller than a critical value.
This regime has not been considered herein.

The numerical investigation of the tachyonic modes and the inclusion of
back reactions promise further
insight into the decay of CP odd metastable states and its realization will
be reported elsewhere.

%%%%%%%%%%%%%%%%%%%%%%%%%%%%%%%%%%%%%%%%%%%%%%%%%%%%%%%%%%%%%%%%%%%%%%%%%%%%%%
\section*{Acknowledgement}
%%%%%%%%%%%%%%%%%%%%%%%%%%%%%%%%%%%%%%%%%%%%%%%%%%%%%%%%%%%%%%%%%%%%%%%%%%%%%%
We thank R. Alkofer and C.D. Roberts  for helpful discussions.
One of us (F.M.S.) acknowledges financial support provided by the DAAD
(Deutscher Akademischer Austauschdienst) allowing him to visit
the Universities  of Rostock and T\"ubingen. 
This work was supported by Deutsche
Forschungsgemeinschaft under project number SCHM 1342/3-1 and AL 279/3-2.
%%%%%%%%%%%%%%%%%%%%%%%%%%%%%%%%%%%%%%%%%%%%%%%%%%%%%%%%%%%%%%%%%%%%%%%%%%%%%%
\appendix
\section{In-field quantization}
%%%%%%%%%%%%%%%%%%%%%%%%%%%%%%%%%%%%%%%%%%%%%%%%%%%%%%%%%%%%%%%%%%%%%%%%%%%%%%
The in-field is a solution of the equation
\begin{equation}
(\Box + m_0^2) {\eta}_{in} =0.
\label{eq: devet}
\end{equation}
The in-field operators fulfill periodic boundary conditions and are
expanded in Fourier modes 
\begin{eqnarray}
{\eta}_{in}(\vec{x},t)& =& \frac{1}{\sqrt{V}} \sum_{\vec{k}} e^{i\vec{k}
\vec{x}} {\eta}_{in}(\vec{k},t),
\label{eq: deset}\\
{\pi}_{in}(\vec{x},t)& =& \frac{1}{\sqrt{V}} \sum_{\vec{k}}
e^{-i\vec{k} \vec{x}} {\pi}_{in}(\vec{k},t),
\end{eqnarray}
where the summation is over discrete momenta $\vec{k} =
\frac{2{\pi}}{L} \vec{n}$, $(n_1,n_2,n_3)$ and 
\begin{eqnarray}
{\eta}_{in}(\vec{k},t)& =& {\Gamma}_{\vec{k}}^0(t) a_{in}(\vec{k})
+ {\Gamma}_{\vec{k}}^{0,{\star}}(t) a_{in}^{\dagger}(-\vec{k}).
\label{eq: odinodin}\\\label{eq: odinshest}
{\pi}_{in}(\vec{k},t) &=&
\dot{\eta}_{in}^{\dagger}(\vec{k},t)\\\nonumber
&=&-i {\omega}_k^0 \Big[ {\Gamma}_{\vec{k}}^0(t) a_{in}(-\vec{k})
- {\Gamma}_{\vec{k}}^{0,{\star}}(t) a_{in}^{\dagger}(\vec{k})
\Big].
\end{eqnarray}
The time-independent creation and annihilation operators obey
the commutation relations
\begin{equation}
[a_{in}(\vec{k}), a_{in}^\dagger (\vec{k}^\prime)]_- =
\delta_{\vec{k},\vec{k}^\prime},
\label{eq: odindva}
\end{equation}
all other commutators vanish.
The function ${\Gamma}_{\vec{k}}^0(t)$ is given by
\begin{equation}
{\Gamma}_{\vec{k}}^0(t) = \frac{1}{\sqrt{2{\omega}_k^0}}
\exp\{-i{\omega}_k^0 t\}
\label{eq: odintri}
\end{equation}
with ${\omega}_k^0 \equiv \sqrt{\vec{k}^2 + m_0^2}$.
Since the field ${\eta}(\vec{x},t)$ is real, we have
${\eta}_{in}^{\dagger} (\vec{k},t) = {\eta}_{in}(-\vec{k},t),$
and
${\pi}_{in}^{\dagger}(\vec{k},t) = {\pi}_{in}(-\vec{k},t).$
The vacuum state $|0;in \rangle \equiv |0 \rangle$ is defined 
as vanishing under the action of
the annihilation operators 
$a_{in}(\vec{k}) | 0 \rangle =0.$

%%%%%%%%%%%%%%%%%%%%%%%%%%%%%%%%%%%%%%%%%%%%%%%%%%%%%%%%%%
\section{Numerical realization}
%%%%%%%%%%%%%%%%%%%%%%%%%%%%%%%%%%%%%%%%%%%%%%%%%%%%%%%%%%
The evolution of ${\phi}$ in the decay is governed
by equation (\ref{eq: tricet}). Introducing the dimensionless
vacuum mean field ${\phi}/f$ and the dimensionless
time variable ${\tau} \equiv {\mu}t$, we rewrite
(\ref{eq: tricet}) as
\begin{equation}
\frac{d^2}{d{\tau}^2} \Big( \frac{{\phi}(\tau)}{f} \Big)
+ \sin\Big( \frac{{\phi}(\tau)}{f} \Big)
+ \frac{a}{{\mu}^2} \Big( \frac{{\phi}(\tau)}{f} \Big)=0,
\label{eq: semsem}
\end{equation}
with the one parameter $a/{{\mu}^2}$ characterizing the solution.

Note that for small  $a/{{\mu}^2} \approx 0$ one can replace
(\ref{eq: semsem}) by the Sine-
Gordon equation 
\begin{equation}
\frac{d^2}{d{\tau}^2} \Big( \frac{{\phi}_0(\tau)}{f} \Big)
+ \sin\Big( \frac{{\phi}_0(\tau)}{f} \Big) =0,
\label{eq: semvosem}
\end{equation}
the subscript $(0)$ in ${\phi}(\tau)$ indicates
the zero value of $a/{{\mu}^2}$. The solution of Eq. (\ref{eq:
  semvosem}) is a 
Jacobian elliptic function. Herein we do not make this approximation
and solve
(\ref{eq: semsem}) numerically for  nonzero values
of $a/{{\mu}^2}$.

For the numerical study we introduce  dimensionless variables for 
the kinetic equations  and obtain
\begin{eqnarray}\nonumber
\frac{d}{d{\tau}} {\cal N}_{+}(\vec{p},{\tau}) =&&
\frac{\dot{\bar{\omega}}_p}{2\bar{\omega}_p}
\int_0^{\tau} d{\tau}^{\prime}  \frac{\dot{\bar{\omega}}_p}
{\bar{\omega}_p}({\tau}^{\prime}) \Big( 1+2{\cal N}_{+}
(\vec{p},{\tau}^{\prime}) \Big)\times\\
&&\cos[2{\Theta}_p^{ad}(\tau) - 2{\Theta}_p^{ad}({\tau}^{\prime})]
\label{eq: semdevet}
\end{eqnarray}
where  the dimensionless frequency is
\begin{equation}
\bar{\omega}_p^2 \equiv \frac{1}{{\mu}^2} {\omega}_k^2
=
{\vec{p}}^{~2} + \cos\Big( \frac{{\phi}(\tau)}{f} \Big) +
\Big( \frac{a}{{\mu}^2} {\Big)}\,.
\label{eq: vosemodin}
\end{equation}
with ${\vec{p}}^{~2} \equiv ({\vec{k}^2}/{{\mu}^2})$.

Eq. (\ref{eq: semdevet}) is an integro-differential equation. It can be
re-expressed by introducing
\begin{eqnarray}\nonumber
u(\vec{p},{\tau}) &\equiv& \int_0^{\tau} d{\tau}^{\prime}
\frac{\dot{\bar{\omega}}_p}{\bar{\omega}_p}({\tau}^{\prime})
\Big( 1 + 2{\cal N}_{+}(\vec{p},{\tau}^{\prime}) \Big)\times\\
&& \sin[2{\Theta}_p^{ad}({\tau}) - 2{\Theta}_p^{ad}({\tau}^{\prime})],
\label{eq: vosemcet}\\
v(\vec{p},{\tau}) &\equiv& \int_0^{\tau} d{\tau}^{\prime}
\frac{\dot{\bar{\omega}}_p}{\bar{\omega}_p}({\tau}^{\prime})
\Big( 1 + 2{\cal N}_{+}(\vec{p},{\tau}^{\prime}) \Big)\times\\
&& \cos[2{\Theta}_p^{ad}({\tau}) - 2{\Theta}_p^{ad}({\tau}^{\prime})],
\label{eq: vosempet}
\end{eqnarray}
with the initial conditions $u(\vec{p},0)=v(\vec{p},0)=0$, in which case
we have
\begin{eqnarray}
\frac{d}{d{\tau}} {\cal N}_{+}(\vec{p},{\tau})& =&
\frac{\dot{\bar{\omega}}_p}{2\bar{\omega}_p} v(\vec{p},{\tau}),
\label{eq: vosemshest}\\
\frac{d}{d{\tau}} v(\vec{p},{\tau})& =& \frac{\dot{\bar{\omega}}_p}
{\bar{\omega}_p} \Big( 1 + 2{\cal N}_{+}(\vec{p},{\tau}) \Big)
- 2\bar{\omega}_p u(\vec{p},{\tau}),
\label{eq: vosemsem}\\
\frac{d}{d{\tau}} u(\vec{p},{\tau})& =& 2\bar{\omega}_p v(\vec{p},{\tau}).
\label{eq: vosemvosem}
\end{eqnarray}


\begin{thebibliography}{99}
\bibitem{QM2000}
Proceedings: QUARK MATTER '99: Proceedings. Edited by L. Riccati,
M. Masera, E. Vercellin. Amsterdam, The Netherlands, North-Holland,
1999. (Nuclear Physics A, Vol. A661, December 1999).
\bibitem{Sauter}
F.~Sauter,
%``\"Uber das Verhalten eines Elektrons im homogenen elektrischen Feld nach 
%der relativistischen Theorie Diracs,''
Z.\ Phys.\ {\bf 69} (1931) 742;
%%CITATION = ZEPYA,69,742;%%
W.~Heisenberg and H.~Euler,
%``Consequences Of Dirac's Theory Of Positrons,''
Z.\ Phys.\ {\bf 98} (1936) 714;
%%CITATION = ZEPYA,98,714;%%
J.~Schwinger,
%``On gauge invariance and vacuum polarization,''
Phys.\ Rev.\ {\bf 82} (1951) 664.
%%CITATION = PHRVA,82,664;%%
\bibitem{tesla}
{\em TESLA -- The Superconductiong Electron Positron Linear Collider
with an Integrated X-Ray Laser Laboratory}, 
Technical Design Report, DESY 2001-011, ECFA 2001-209, TESLA-Report 2001-23,
TESLA-FEL 2001-05;
\bibitem{xfel} A.~Ringwald,
%``Pair production from vacuum at the focus of an X-ray free electron  laser,''
Phys.\ Lett.\  {\bf B 510} (2001) 107;
%%CITATION = HEP-PH 0103185;%%
R.~Alkofer, et al.,
%``Pair Creation and an X-ray Free Electron Laser,''
nucl-th/0108046, Phys.\ Rev.\ Lett. in press.
%%CITATION = NUCL-TH 0108046;%%
\bibitem{Damour}
T.~Damour and R.~Ruffini,
%``Black Hole Evaporation In The Klein-Sauter-Heisenberg-Euler Formalism,''
Phys.\ Rev.\ {\bf D 14} (1976) 332.
%%CITATION = PHRVA,D14,332;%%
\bibitem{Parker}
L.~Parker,
%``Quantized Fields And Particle Creation In Expanding Universes. 1,''
Phys.\ Rev.\ {\bf 183} (1969) 1057.
%%CITATION = PHRVA,183,1057;%%
\bibitem{Casher}
A.~Casher, H.~Neuberger and S.~Nussinov,
%``Chromoelectric Flux Tube Model Of Particle Production,''
Phys.\ Rev.\ {\bf D 20} (1979) 179;
%%CITATION = PHRVA,D20,179;%%
B.~Andersson, G.~Gustafson, G.~Ingelman and T.~Sj\"ostrand,
%``Parton Fragmentation And String Dynamics,''
Phys.\ Rept.\ {\bf 97} (1983) 31;
%%CITATION = PRPLC,97,31;%%
T.~S.~Biro, H.~B.~Nielsen and J.~Knoll,
%``Color Rope Model For Extreme Relativistic Heavy Ion Collisions,''
Nucl.\ Phys.\ {\bf B 245} (1984) 449.
%%CITATION = NUPHA,B245,449;%%
\bibitem{kme} 
S.A. Smolyansky et al., hep-ph-9712377;
%%CITATION = HEP-PH 9712377;%%
S.M. Schmidt et al., Int.\ J.Mod.\ Phys. {\bf E 7} (1998) 709;
%%CITATION = HEP-PH 9809227;%%
Y. Kluger, E. Mottola, and J.M. Eisenberg,
Phys.\ Rev. {\bf D 58} (1998) 125015.
%%CITATION = HEP-PH 9803372;%%
\bibitem{rau} 
J. Rau and B. M\"uller, Phys.\ Rep. {\bf 272} (1996) 1.
%%CITATION = PRPLC,272,1;%%
\bibitem{sms} S.M. Schmidt et al., Phys.\ Rev. {\bf D 59} (1999)
094005; 
%%CITATION = HEP-PH 9810452;%%
S.~M.~Schmidt, A.~V.~Prozorkevich and S.~A.~Smolyansky,
%``Creation of boson and fermion pairs in strong fields,''
hep-ph/9809233;
%%CITATION = HEP-PH 9809233;%%
J.C.R. Bloch, C.D. Roberts, and
S.M. Schmidt, Phys.\ Rev. {\bf D 61} (2000) 117502.
%%CITATION = NUCL-TH 9910073;%%
\bibitem{kkaj} 
K. Kajantie and T. Matsui, Phys.\ Lett.
{\bf B 146} (1985) 373; 
%%CITATION = PHLTA,B164,373;%%
G. Gatoff, A.K. Kerman and
T. Matsui, Phys.\ Rev. {\bf D 36} (1987) 114.
%%CITATION = PHRVA,D36,114;%%
\bibitem{jmeis} J.M. Eisenberg and G. K\"albermann,
Phys.\ Rev. {\bf D 37} (1988) 1197;
%%CITATION = PHRVA,D37,1197;%%
 Y. Kluger et al., Phys.\ Rev.\ Lett. {\bf 67} (1991) 2427; 
%%CITATION = PRLTA,67,2427;%%
Phys.\ Rev. {\bf D 45} (1992) 4659; 
%%CITATION = PHRVA,D45,4659;%%
F. Cooper et al., Phys.\ Rev. {\bf D 48} (1993) 190; 
%%CITATION = HEP-PH 9212206;%%
\bibitem{nayak}
R.S.~Bhalerao and G.C.~Nayak,
%``Production and equilibration of quark-gluon plasma at RHIC and LHC with  minijets,''
Phys.\ Rev.\ C {\bf 61} (2000) 054907;%%CITATION = HEP-PH 9907322;%%
Q.~Wang, C.~Kao, G.~C.~Nayak, H.~Stoecker and W.~Greiner,
%``Color current induced by gluons in background field method of QCD,''
hep-th/0009076;
%%CITATION = HEP-TH 0009076;%%
K.~Bajan and W.~Florkowski,
%``Boost-invariant particle production in transport equations,''
hep-ph/0107244.
%%CITATION = HEP-PH 0107244;%%
\bibitem{bloch}
J.~Bloch et al.,
%``Pair creation: Back-reactions and damping,''
Phys.\ Rev. {\bf D 60} (1999) 116011,
%%CITATION = NUCL-TH 9907027;%%
A.V.~Prozorkevich et al.,
%``Pair creation and plasma oscillations,''
nucl-th/0012039;
%%CITATION = NUCL-TH 0012039;%%
D.V.~Vinnik et al.,
%``Plasma production and thermalisation in a strong field,''
nucl-th/0103073, Eur. Phys. J. {\bf C} in press.
%%CITATION = NUCL-TH 0103073;%%
\bibitem{review}
C.D. Roberts and S.M. Schmidt, Prog.\ Part.\ Nucl.\ Phys.
{\bf 45} (2000) S1.
%%CITATION = NUCL-TH 0005064;%%
\bibitem{edwin}
E.~Laermann,
%``Finite-temperature QCD on the lattice,''
Phys.\ Part.\ Nucl.\  {\bf 30} (1999) 304
[Fiz.\ Elem.\ Chast.\ Atom.\ Yadra {\bf 30} (1999) 720].
%%CITATION = PPNUE,30,304;%%
\bibitem{alk3}
R.~Alkofer and L.~von Smekal, Phys.\ Rept. {\bf 353} (2001) 281.
%%CITATION = HEP-PH 0007355;%%
\bibitem{trento}
ECT* International Workshop on Understanding Deconfinement in QCD,
   Trento, Italy, 1-13 Mar 1999.
   UNDERSTANDING DECONFINEMENT IN QCD: Proceedings.  Edited
   by David Blaschke, Frithjof Karsch, Craig D. Roberts. World
   Scientific, 2000.
%
\bibitem{njl}
S.~P.~Klevansky,
%``The Nambu-Jona-Lasinio model of quantum chromodynamics,''
Rev.\ Mod.\ Phys.\  {\bf 64} (1992) 649.
%%CITATION = RMPHA,64,649;%%
\bibitem{Bender}
A.~Bender  et al.,
%``Continuum study of deconfinement at finite temperature,''
Phys.\ Rev.\ Lett.\  {\bf 77} (1996) 3724;
%%CITATION = NUCL-TH 9606006;%%
C.~D.~Roberts,
%``Nonperturbative effects in QCD at finite temperature and density,''
Phys.\ Part.\ Nucl.\  {\bf 30} (1999) 223
[Fiz.\ Elem.\ Chast.\ Atom.\ Yadra {\bf 30} (1999) 537].
\bibitem{dkhar}
D. Kharzeev, R.D. Pisarski, and M.H.C. Tytgat,
Phys.\ Rev.\ Lett. {\bf 81} (1998) 512;
%%CITATION = HEP-PH 9804221;%%
D. Kharzeev and R.D. Pisarski,
Phys.\ Rev. {\bf D 61} (2000) 111901;
%%CITATION = HEP-PH 9906401;%%
D. Kharzeev, A. Krasnitz, and R. Venugopalan, hep-ph/0109253.
%%CITATION = HEP-PH 0109253;%%
\bibitem{jkap} 
J. Kapusta, D. Kharzeev, and L. McLerran,
Phys.\ Rev. {\bf D 53} (1996) 5028; 
%%CITATION = HEP-PH 9507343;%%
Z. Huang and X.-N. Wang,
Phys.\ Rev. {\bf D 53} (1996) 5034.
%%CITATION = HEP-PH 9507395;%%
\bibitem{alk1}
R.~Alkofer, P.~A.~Amundsen and H.~Reinhardt,
%``Temperature Dependence Of Eta And Eta-Prime Production In Heavy Ion Collisions,''
Phys.\ Lett. {\bf B 218} (1989) 75.
%%CITATION = PHLTA,B218,75;%%
\bibitem{dahr} 
D. Ahrensmeier, R. Baier, and M. Dirks,
Phys.\ Lett. {\bf B 484} (2000) 58.
%%CITATION = HEP-PH 0005051;%%
\bibitem{gvene} 
G. Veneziano, Nucl.\ Phys. {\bf B 159} (1979) 213;
%%CITATION = NUPHA,B159,213;%%
P. Di Vecchia and G. Veneziano, Nucl.\ Phys. {\bf B 171} (1980) 253;
%%CITATION = NUPHA,B171,253;%%
P. Di Vecchia et al., Nucl.\ Phys. {\bf B 181} (1981) 318;
%%CITATION = NUPHA,B181,318;%%
E. Witten, Nucl.\ Phys. {\bf B 156} (1979) 269;
Annals Phys. {\bf 128} (1980) 363; 
%%CITATION = APNYA,128,363;%%
Phys.\ Rev.\ Lett. {\bf 81} (1998) 2862.
%%CITATION = HEP-TH 9807109;%%
\bibitem{alk}
H.~Reinhardt and R.~Alkofer,
%``Instanton Induced Flavor Mixing In Mesons,''
Phys.\ Lett.\  {\bf B 207}  (1988) 482;
%%CITATION = PHLTA,B207,482;%%
\bibitem{alk2}
R.~Alkofer and I.~Zahed,
%``The Pseudoscalar Nonet In Generalized Njl Models,''
Phys.\ Lett.\  {\bf B 238} (1990) 149.
%%CITATION = PHLTA,B238,149;%%
\bibitem{mesons}
P.~Maris, C.~D.~Roberts and S.~M.~Schmidt,
%``Chemical potential dependence of pi and rho properties,''
Phys.\ Rev.\  {\bf C 57} (1998) 2821;
%%CITATION = NUCL-TH 9801059;%%
P.~Maris et al.,
%``T-dependence of pseudoscalar and scalar correlations,''
Phys.\ Rev.\  {\bf C 63} (2001) 025202;
%%CITATION = NUCL-TH 0001064;%%
D.~Blaschke et al.,
%``Finite T meson correlations and quark deconfinement,''
Int.\ J.\ Mod.\ Phys.\ {\bf A 16} (2001) 2267.
%%CITATION = NUCL-TH 0002024;%%
\bibitem{land}
 L. Landau and E. Lifshitz, {\it Mechanics}
(Pergamon, Oxford, 1960); V. Arnold, {\it Mathematical Methods of
Classical Mechanics} (Springer, New York, 1978).
\bibitem{tras} 
J. Traschen and R. Brandenberger, Phys.\ Rev. {\bf D 42}
(1990) 2491; 
%%CITATION = PHRVA,D42,2491;%%
R.H. Brandenberger, hep-ph-0102183.
%%CITATION = HEP-PH 0102183;%%
\bibitem{kirz} 
D.A. Kirzhnits and A.D. Linde, Phys.\ Lett. {\bf B 42}
(1972) 471; 
%%CITATION = PHLTA,B42,471;%%
Ann.\ Phys. (NY) {\bf 101} (1976) 195; 
%%CITATION = APNYA,101,195;%%
S. Weinberg,
Phys.\ Rev. {\bf D 9} (1974) 3357; 
%%CITATION = PHRVA,D9,3357;%%
L. Dolan and R. Jackiw, Phys.\ Rev.
{\bf D 9} (1974) 3320.
%%CITATION = PHRVA,D9,3320;%%
\bibitem{anse} 
A.A. Anselm and M.G. Ryskin, Phys.\ Lett. {\bf B 266}
(1991) 482; 
%%CITATION = PHLTA,B266,482;%%
K. Rajagopal and F. Wilczek, Nucl.\ Phys. {\bf B 399} (1993)
395;
%%CITATION = HEP-PH 9210253;%% 
{\bf B 404} (1993) 577; 
%%CITATION = HEP-PH 9303281;%%
D. Boyanovsky, H.J. de Vega, and
R. Holman, Phys.\ Rev. {\bf D 51} (1995) 734.
%%CITATION = HEP-PH 9401308;%%
\bibitem{linde} 
%A. Linde, Phys.Lett. {\bf B259} (1991) 38; 
%%CITATION = PHLTA,B259,38;%%
%Phys.Rev. {\bf D49} (1994) 748; 
%%CITATION = ASTRO-PH 9307002;%%
%J. Garcia-Bellido and A. Linde, Phys.Rev.
%{\bf D57} (1998) 6065; 
G. Felder et al., Phys.\ Rev.\ Lett. {\bf 87}
(2001) 011601.
%%CITATION = HEP-PH 0012142;%%
% L. Kofman, hep-ph-0012298; 
%%CITATION = HEP-PH 0012298;%%
\bibitem{felder}
G. Felder, L. Kofman, 
and A. Linde, hep-th-0106179.
%%CITATION = HEP-TH 0106179;%%


\end{thebibliography}
\end{document}